\documentclass{eas}
\usepackage{graphicx}
\begin{document}

\runningtitle{Magnetism in Herbig Ae/Be stars}
\title{Magnetism in Herbig Ae/Be stars and the link to the Ap/Bp stars} 
\author{E. Alecian}\address{Royal Military College, CANADA}
\author{C. Catala}\address{(LESIA (Observatoire de Paris), FRANCE}
\author{G.A. Wade$^1$}
\author{S. Bagnulo}\address{Armagh Observatory, U.K.}
\author{T. Boehm}\address{LATT (Observatoire Midi-Pyr\'en\'ees), FRANCE}
\author{J.-C. Bouret}\address{Laboratoire d'astrophysique de Marseille, FRANCE}
\author{J.-F. Donati$^4$}
\author{C. Folsom$^3$}
\author{J. Grunhut$^1$}
\author{J.D. Landstreet}\address{University of Western Ontario, CANADA}
\author{S.C. Marsden}\address{Anglo-Australian Observatory, AUSTRALIA}
\author{P. Petit$^4$}
\author{J. Ramirez$^2$}
\author{J. Silvester$^1$}
\begin{abstract}
Among the A/B stars, about 5\% host large-scale organised magnetic fields. These magnetic stars show also abundance anomalies in their spectra, and are therefore called the magnetic Ap/Bp stars. Most of these stars are also slow rotators compared to the normal A and B stars.
Today, one of the greatest challenges concerning the Ap/Bp stars is to understand the origin of their slow rotation and their magnetic fields. The favoured hypothesis for the latter is that the fields are fosils, which implies that the magnetic fields subsist throughout the different evolutionary phases, and in particular during the pre-main sequence phase. The existence of magnetic fields at the pre-main sequence phase is also required to explain the slow rotation of Ap/Bp stars. 

During the last 3 years we performed a spectropolarimetric survey of the Herbig Ae/Be stars in the field and in young clusters, in order to investigate their magnetism and rotation. These investigations have resulted in the detection and/or confirmation of magnetic fields in 8 Herbig Ae/Be stars, ranging in mass from 2 to nearly 15 solar masses. In this paper I will present the results of our survey, as well as their implications for the origin and evolution of the magnetic fields and rotation of the A and B stars.
\end{abstract}
\maketitle

%
\section{Introduction}
\subsection{Magnetism and rotation in the main sequence A and B stars}

Between about 1.5 and 10 M$_{\odot}$, at spectral types A and B, about 5\% of main sequence (MS) stars have magnetic fields with characteristic strengths of about 1 kG. Such stars also show important chemical peculiarities and are thus usually called the magnetic chemically peculiar Ap/Bp stars. Two hypotheses have been proposed to explain the origin of their magnetic fields: the core dynamo hypothesis and the fossil field hypothesis. The core dynamo hypothesis assumes that a magnetic field is generated in the convective core of the main sequence (MS) Ap/Bp stars, and then diffuses towards the surface. On the contrary, the fossil field hypothesis assumes that the stellar magnetic fields are relics from the field present in the parental interstellar cloud or possibly generated early in their evolution, by a dynamo that has ceased to operate. It implies that magnetic fields can (at least partially) survive the violent phenomena accompanying the birth of stars, and can also remain throughout their evolution and until at least the end of the MS, without significant regeneration.

According to the fossil field model, some pre-main sequence (PMS) stars of intermediate mass should be magnetic. Many such PMS stars are identifiable observationally as Herbig Ae/Be (HAeBe) stars. However no magnetic field was observed up to recently in these stars (except in HD~104237, Donati et al., 1997). {\it Can we obtain observational evidence of the presence of magnetic fields 
in PMS A/B stars, as predicted by the fossil field hypothesis? If some HAeBe stars are discovered to have magnetic fields, is the fraction of magnetic to non-magnetic HAeBe stars the same as the fraction for main sequence stars? Are the magnetic fields in HAeBe stars strong enough and of appropriate geometry to explain those of the Ap/Bp stars?}

Chemical peculiarities and magnetism are not the only characteristic properties observed in the Ap/Bp stars. Most magnetic MS stars have rotation periods (typically of a few days) that are several times longer than the rotation periods of non-magnetic MS stars (a few hours to one day). It is usually believed that magnetic braking, in particular during PMS evolution when the star can exchange angular momentum with its massive accretion disk, is responsible for this slow rotation (St{\c e}pie{\'n}, 2000). An alternative to this picture involves a rapid dissipation of the magnetic field during the early stages of PMS evolution for the fastest rotators, due to strong turbulence induced by rotational shear developed under the stellar surface, as occurs in solar-type stars (e.g. Ligni\`eres, 1996). In this scenario, only slow rotators could retain their initial magnetic fields, and evolve as magnetic stars to the main sequence. So another question to be addressed is the following: {\it does the magnetic field control the rotation of the star, or does the rotation of the star control the magnetic field?}

To answer to these questions we adopt two strategies. The first consists of observing the brightest "field" HAeBe stars in order to detect magnetic fields, to measure rotation velocities and finally to make a statistical study to be compared to the statistical properties of magnetism and rotation of the main sequence A/B stars. The second strategy consists of observing HAeBe stars in young open clusters and associations. Observing stars in a single cluster or association will provide us with stars of approximately the same age and initial conditions (chemical composition, angular momentum, environment). Then, by selecting members of open clusters and associations of various ages, we hope to be able to disentangle "evolutionary" effects from "initial conditions" effects. Thanks to both studies we will therefore improve our understanding  of the evolution of the magnetic field during the pre-main sequence phase, and its impact on the evolution of the stars.

%
\subsection{Evolution of intermediate-mass stars}

The MS A and B stars are intermediate-mass (1.5 to 15~M$_{\odot}$) stars. The main difference between MS intermediate mass stars and MS low-mass star is their internal structure. A low-mass star has a radiative core and a convective envelope, while A and B stars have a convective core and a radiative envelope. In low-mass stars, magnetic fields are generated by dynamos in their convective envelope. We understand therefore the necessity to involve a different origin for the magnetic field of the intermediate mass stars.

\begin{figure}[t!]
\centering
\includegraphics[width=6cm]{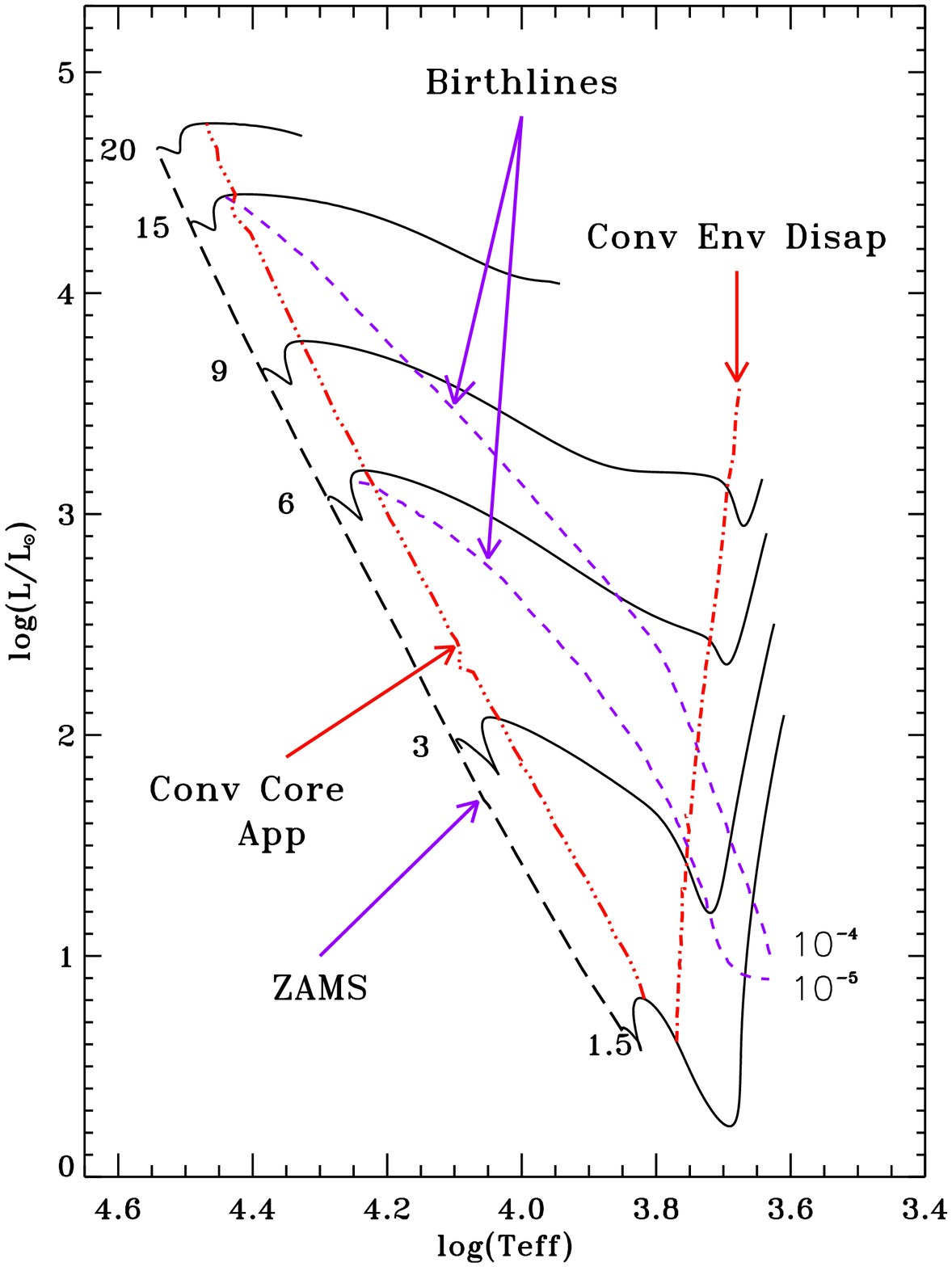}
\hfill
\includegraphics[width=6cm]{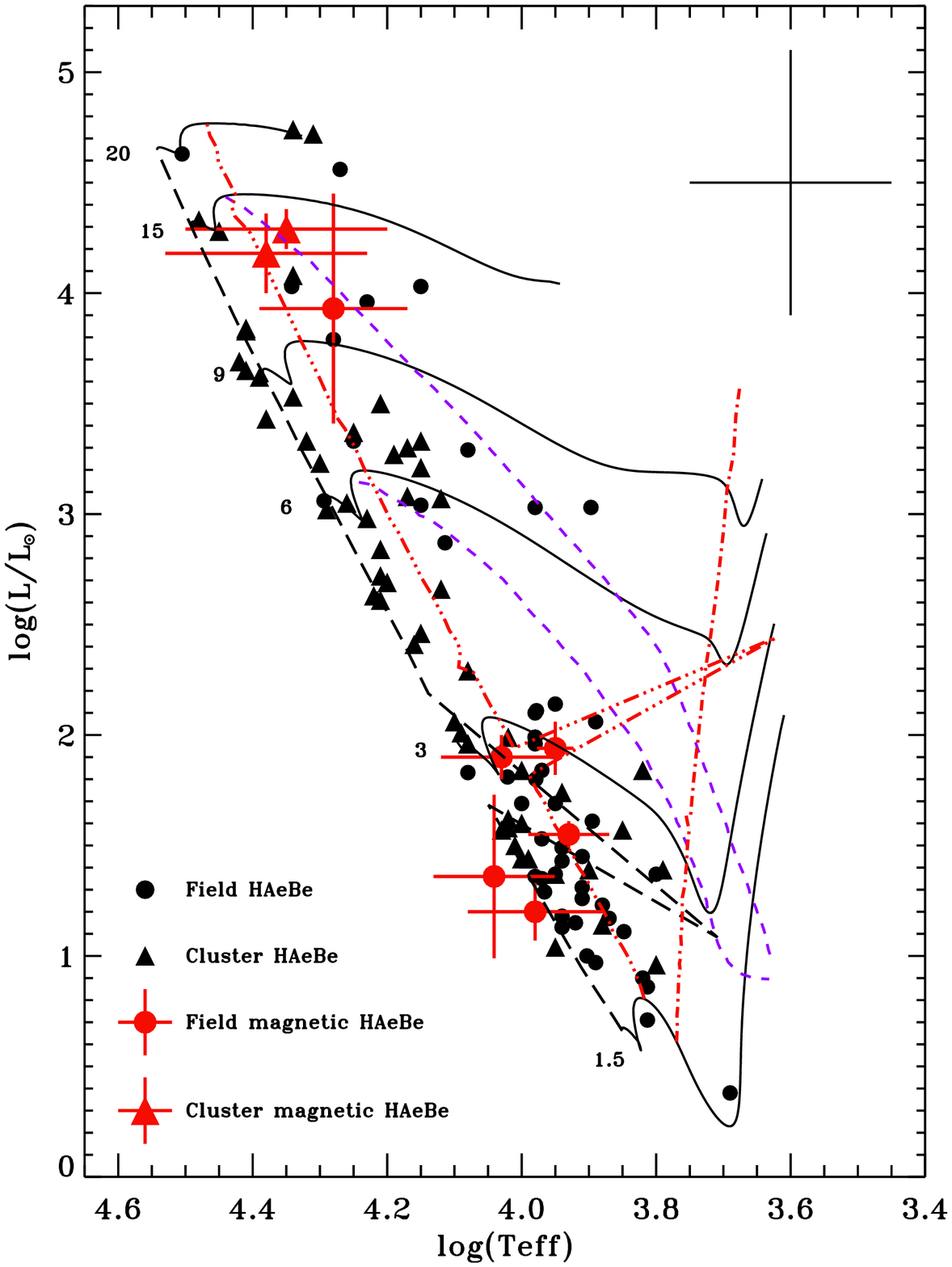}
\caption{{\it Left: }  PMS evolutionary tracks (full black line), convective envelope disappearance (red dot-dashed line) and convective core appearance (red dot-dot-dot-dashed line), calculated with CESAM (Morel 1997). The black long-dashed line is the ZAMS and the blue small-dashed lines are the birthlines for 10$^{-5}$ and 10$^{-4}$ $M_{\odot}$.yr$^{-1}$ mass accretion rates (Palla \& Stahler 1993). {\it Right: } Field (circles) and cluster (triangles) HAeBe stars. The large cross in the upper-right of the panel is the mean error bars in luminosity and temperature of the undetected stars.}
\label{fig:hr}
\end{figure}

During their pre-main sequence phase the intermediate and low mass stars follow a different evolution, which is fundamental in understanding the origin of magnetic fields. In Fig. \ref{fig:hr} (left) are plotted the PMS evolutionary tracks (full black lines) computed with the CESAM code (Morel 1997), for different masses (from 1.5 to 20 M$_{\odot}$). The PMS evolution starts with a fully convective phase on the Hayashi track (the vertical part). During many years, the fossil field hypothesis was considered to be unlikely because magnetic field relics would have been dissipated during this convective phase. However Palla \& Stahler (1993) modelled proto-stellar evolution until the star becomes optically visible by clearing away its opaque envelope. They defined the birthline as the locus in the HR diagram of these newly formed stars, and they computed it using two different mass accretion rates during the protostellar phase : 10$^{-5}$~M$_{\odot}$.yr$^{-1}$ and 10$^{-4}$~M$_{\odot}$.yr$^{-1}$ . Both birthlines are plotted in Fig. \ref{fig:hr} (left) in small dashed-lines. The zero-age main-sequence (ZAMS) computed with CESAM is also plotted in long dashed-line. The PMS lifetime of a star is therefore from the birthline to the ZAMS. It implies that the Hayashi phase of intermediate mass stars is considerably reduced for stars between 1.5 $M_{\odot}$ and 3 $M_{\odot}$ and disappears totally above 3~$M_{\odot}$. Fossil magnetic fields can therefore survive during the PMS evolution of intermediate mass stars.

In Fig. \ref{fig:hr} are also plotted the convective envelope disappearance and the convective core appearance lines, calculated with CESAM. During the PMS phase stars experience three different internal structures: either a radiative core and a convective envelope, or a totally radiative interior, or a convective core and a radiative envelope. The PMS stars of intermediate mass stars are therefore very suitable laboratories to test both hypotheses of magnetic field origin.

%
\subsection{The Herbig Ae/Be stars}

The Herbig Ae/Be stars are intermediate mass stars displaying emission lines in their spectrum. They are distinguished from the classical Be stars by their abnormal extinction law and association with nebulae. Most of them are still in the pre-main sequence phase, and therefore the evolutionary progenitors of the MS A and B stars. 

HAeBe stars display many observational phenomena often associated with magnetic activity. First, highly ionised lines are observed in the spectra of some stars (e.g. Bouret et. al., 1997), and X-ray emission has been detected coming from the environments of HAeBe stars (e.g. Hamaguchi et al., 2005). In active cool stars, many of these phenomena are produced in hot chromospheres or coronae. Some authors have also mentioned rotational modulation of resonance lines which they speculate may be due to rotation modulation of winds structured by magnetic field (e.g. Catala et al. 1989). 

In the literature we find many clues to the presence of circumstellar disks around these stars, from spectroscopic and photometric data (e.g. Vink et al. 2002). Recently, using coronographic and interferometric data, some authors have also found direct evidence of circumstellar disks around these stars (e.g. Eisner et al. 2003). These disks show similar properties to the disks of their low mass counterparts (Natta et al. 2001), the T Tauri stars, whose emission lines have been explained by magnetospheric accretion models (Muzerolle et al. 2001). Finally Muzerolle et al. (2004) have sucessfully applied their magnetospheric accretion model to one HAeBe star to explain the emission lines in its spectrum. 

For all these reasons we suspect that the HAeBe stars may host large-scale magnetic fields that should be detectable with current instrumentation. However, many authors have tried to detect such fields without much success (Catala et al. 1989, 1993; Donati et al. 1997; Hubrig et al. 2004; Wade et al. 2007). The previous instruments were limited either by their spectral range, or by their spectral resolution, or by low efficiency. However, in 2005 a new high-resolution spectropolarimeter, ESPaDOnS, was installed at the Canada-France-Hawaii telescope. This powerful new instrument has provided us with the capability to sensitively survey many HAeBe stars in order to investigate rotation and magnetism in the pre-main sequence stars of intermediate mass.

%
\section{Observations}

Our data were obtained using the high resolution spectropolarimeter ESPaDOnS installed on the 3.6 m Canada-France-Hawaii Telescope (CFHT, Donati et al., in preparation; Neiner 2008, these proceedings) during several scientific runs.

We used this instrument in polarimetric mode, measuring Stokes $I$ and $V$ spectra of 65000 resolution. The data were reduced using the "Libre ESpRIT" package especially developed for ESPaDOnS, and installed at the CFHT (Donati et al. 1997; Donati et al. in preparation).

We then applied the Least Squares Deconvolution procedure to all spectra (Donati et al. 1997), in order to increase our signal to noise ratio. This method assumes that all selected lines of the intensity spectrum have a profile of similar shape. Hence, this supposes that all lines are broadened in the same way. We can therefore consider that the observed spectrum is a convolution between a profile (which is the same for all lines) and a mask including all choosen lines of the spectrum. We therefore apply a deconvolution to the observed spectrum using the mask, in order to obtain the average photospheric profiles of Stokes $I$ and $V$. In this procedure, each line is weighted by its signal to noise ratio, its depth in the unbroadened model and its Land\'e factor. For each star we used a mask computed using "extract stellar" line lists obtained from the Vienna Atomic Line Database (VALD)\footnote{http://www.astro.univie.ac.at/$\sim$vald/}, with effective temperatures and $\log g$ suitable for each star (Wade et al. in prep.). We excluded from this mask hydrogen Balmer lines, strong resonance lines, lines whose Land\'e factor is unknown and emission lines. The results of this procedure are the mean Stokes $I$ and Stokes $V$ LSD profiles.

%
\section{Field Herbig Ae/Be stars study}
\subsection{Our sample}



Our sample contains 55 stars which have masses ranging from 1.5 M$_{\odot}$ to 20 M$_{\odot}$, selected in the catalogues of Th\'e et al. (1994) and Vieira et al. (1993) with a visual magnitude brighter than 12. In Fig. \ref{fig:hr} (right, circle) are plotted the stars of our sample in an HR diagram. We observe a variety of internal structure of stars in our sample will allow us to test both the fossil field hypothesis and the core dynamo hypothesis.

%
\subsection{The magnetic field}

\begin{figure}[t]
\centering
\includegraphics[width=6.cm, angle=90]{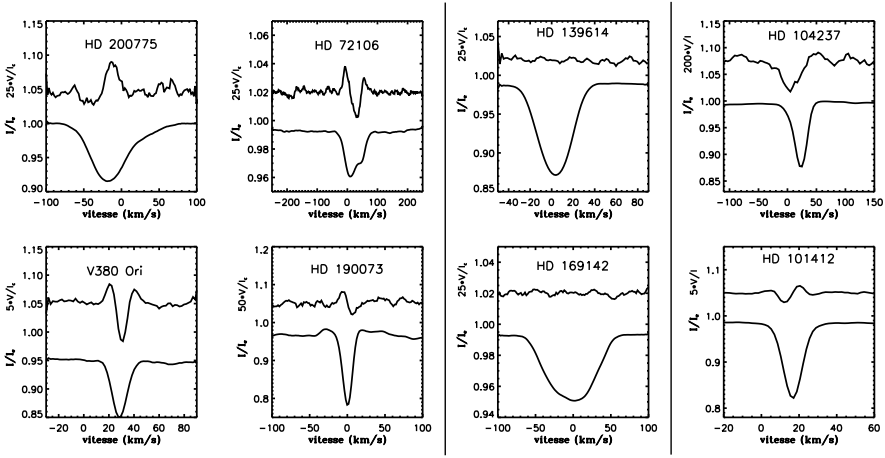}
\caption{Stokes $I$ (bottom) and Stokes $V$ (top) LSD profiles of the 4 magnetic stars observed with ESPaDOnS (left), of 2 undetected stars (middle) and of the 2 magnetic stars observed with SemelPol (right). Note the amplification factor in $V$.}
\label{fig:compil}
\end{figure}

The combination of ESPaDOnS and the LSD method has allowed us to discover four new magnetic HAeBe stars. The Zeeman signatures in the Stokes $V$ profiles of the four magnetic HAeBe stars are plotted in Fig. \ref{fig:compil} on the left. These can be compared to some undetected stars (stars in which we have not detected magnetic fields) in the middle. We observe that these signatures are located at the same position as the absorption line, and are as broad as the intensity profile. They are therefore characteristic of the presence of magnetic fields in the stars. As our sample contains 55 stars, the first conclusion that we draw is that ~7\% of our sample HAeBe stars are magnetic.

In order to characterise these magnetic fields, we monitored the four detected stars during many nights. We recorded variations of the Zeeman signature with time for all stars, except one (HD~190073), which exhibits a constant profile. To model these variations we assume the oblique rotator model described in Landstreet (1970). This model assumes a dipole of intensity $B_{\rm d}$ placed at a distance $d_{\rm dip}$ of the center of the star, on the magnetic axis inside the star, and inclined with an angle $\beta$ to the rotation axis. The rotation axis make an angle $i$ with the line of sight. As the star rotates with a period $P$, the magnetic configuration at the visible surface of the star, and therefore the Stokes $V$ profile, changes with the rotation phase. We calculated on each point of the observable surface of the star a local intensity $I(\theta,\phi)$ profile using a Gaussian of instrumental width placed at the local velocity. Using the weak field approximation (Landi degl'Innocenti \& Landi degl'Innocenti 1973), we calculated the local Stokes $V(\theta,\phi)$ profile as a function of ${\rm d}I(\theta,\phi)/{\rm dv}$ and the local magnetic field projected on the line of sight $b_{\ell}(\theta,\phi)$. We then integrated $I(\theta,\phi)$ and $V(\theta,\phi)$ over the stellar surface, using the linear limb-darkening law with a suitable coefficient value for the star (Claret 2000), in order to get the Stokes $I$ and $V$ profiles. This model is therefore dependent on 6 parameters: the rotation period $P$, the reference time of the rotation phase $T_0$, the magnetic obliquity $\beta$, the inclination of the rotation axis $i$, the dipole intensity $B_{\rm d}$, and the position $d_{\rm dip}$ of the dipole on the magnetic axis with respect to the center of the star.

\begin{figure}[t]
\centering
\includegraphics[width=6.cm, angle=90]{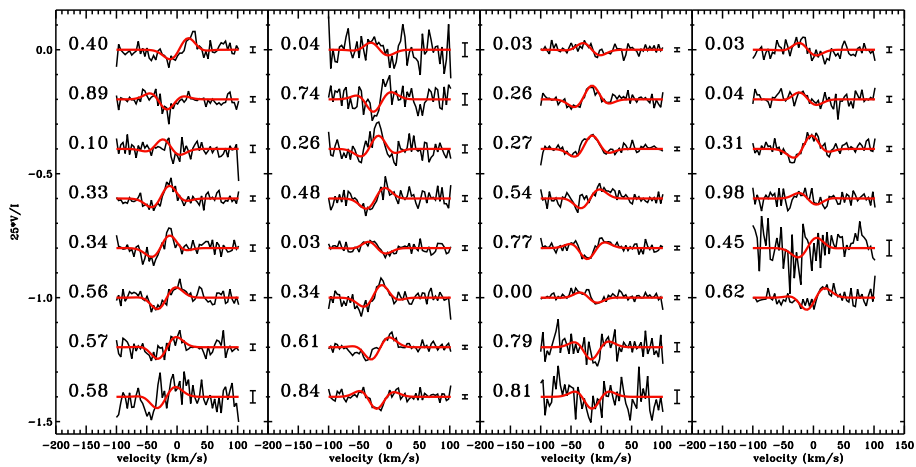}
\caption{Stokes $V$ profiles of HD~200775 superimposed on the synthetic ones, corresponding to our best fit. The rotation phase and the error bars are indicated on the left and on the right of each profile, respectively (Alecian et al. 2008a).}
\label{fig:fitv}
\end{figure}

We fitted the observed Stokes $I$ profiles to the synthetic ones, described above, then we calculated a grid of synthetic Stokes $V$ profiles by varying the 6 parameters. Finally, we fitted all observed $V$ profiles of a single star by a $\chi^2$ minimisation. The best model of HD~200775 is superimposed on the observed profiles in Fig. \ref{fig:fitv}, while the best model parameters are summarised in Table \ref{tab:fitv} for all the stars. In the case of V380 Ori we found two possible models. We do not have yet enough data to choose between them. In all cases the oblique rotator model is sufficient to reproduce our observations.

\begin{table}[t]
\caption{Fundamental, geometrical and magnetic parameters of the magnetic Herbig Ae/Be stars. References : 1 : Alecian et al. (2008a), 2 : Folsom et al. (2008), 3 : Alecian et al., in prep., 4 : Catala et al. (2007).}
\label{tab:fitv}
\centering
\begin{minipage}[t]{16cm}
\begin{tabular}{lcccccccc}
\hline
\footnotesize{Star}   & \footnotesize{S.T}. & \footnotesize{P}    & \footnotesize{$B_{\rm P}$} & \footnotesize{$\beta$ ($^{\circ}$)} & \footnotesize{$i$ ($^{\circ}$)} & \footnotesize{$d_{\rm dip}$} & \footnotesize{$B_{\rm P (ZAMS)}$} & \footnotesize{Ref.} \\
         &         & \footnotesize{(d)}  & \footnotesize{(kG)}             &                                 &                         & \footnotesize{$R_*$}             & \footnotesize{(kG)}                          &        \\
\hline
\footnotesize{HD 200775} & \footnotesize{B2}  & \footnotesize{4.3281}         & \footnotesize{1.0}             & \footnotesize{55}            & \footnotesize{60}            & \footnotesize{0.05} & \footnotesize{3.6}                                             & 1 \\
\footnotesize{HD 72106}   & \footnotesize{A0}  & \footnotesize{0.63995}     & \footnotesize{1.3}          & \footnotesize{60}            & \footnotesize{23}            & \footnotesize{0}    & \footnotesize{1.3}                                             & \footnotesize{2} \\
\footnotesize{V380 ori} 
& \footnotesize{A2}  & \footnotesize{$[7.6,9.8]$} & \footnotesize{1.4}          & \footnotesize{$[90,85]$} & \footnotesize{$[36,49]$} & \footnotesize{0}    & \footnotesize{2.4} & \footnotesize{3} \\
\footnotesize{HD 190073}
& \footnotesize{A2}  &                   & \footnotesize{$[0.1,1]$} & \footnotesize{$[0,90]$}   & \footnotesize{$[0,90]$}   &       & \footnotesize{$[0.4,4]$}                                    & \footnotesize{4} \\
\hline
\end{tabular}
\end{minipage}
\end{table}

This method is very efficient if we observe variations in the Stokes $V$ profiles. However in the case of HD~190073, no such variations are observed. To explain this we propose 3 hypotheses:  either the star is seen pole-on, or the magnetic obliquity is null and the magnetic field is axisymetric, or the rotation period of the star is very long. All these hypotheses are consistent with a dipolar field inside the star, which is further strengthened by the fact that the Zeeman signature is a simple bipolar signature which is stable over more than 2 years.

Finally, we mention two other detections obtained using the SemelPol polarimeter and the UCLES spectrograph at the Anglo-Australian Telescope (AAT) (Semel et al. 1993, Donati et al. 1997). The first is HD~104237, which was discovered to be magnetic by Donati et al. (1997), and which has been recently confirmed (Wade et al., in preparation). The second is HD 101412, discovered to be magnetic by Wade et al. (2007) using FORS1 at VLT, and which has been recently confirmed using SemelPol. The Zeeman signatures are plotted in Fig. \ref{fig:compil} on the right, and are typical signatures of a magnetic field inside the star. We still don't have enough data to characterise their magnetic fields, but the simple signatures are consistent with organised fields.

%
\subsection{The rotation}


In order to study the rotation during the PMS and MS phases, as well as the magnetic braking hypothesis, we measured the projected rotational velocities ($v\sin i$), by performing a Least-Squares fit to the LSD $I$ profiles of each star of our sample. The fitted function is the convolution of a rotation function (for which $v\sin i$ is a free parameter in the fit) and a gaussian whose width is fixed and computed from the spectral resolution and the inferred macroturbulent velocity (Gray 1992). The free parameters of the fitting procedure are the centroids, depths, and $v\sin i$.

In our study, we separated the stars of our sample in two different groups: the magnetic HAeBe stars, and the normal HAeBe stars which contains the non-magnetic and non-multiple stars (as the tidal interactions between both components of a binary system are braking both stars, we prefer to not include them in our study of the rotational evolution during the PMS phase).

\begin{figure}[t!]
\centering
\includegraphics[width=4cm]{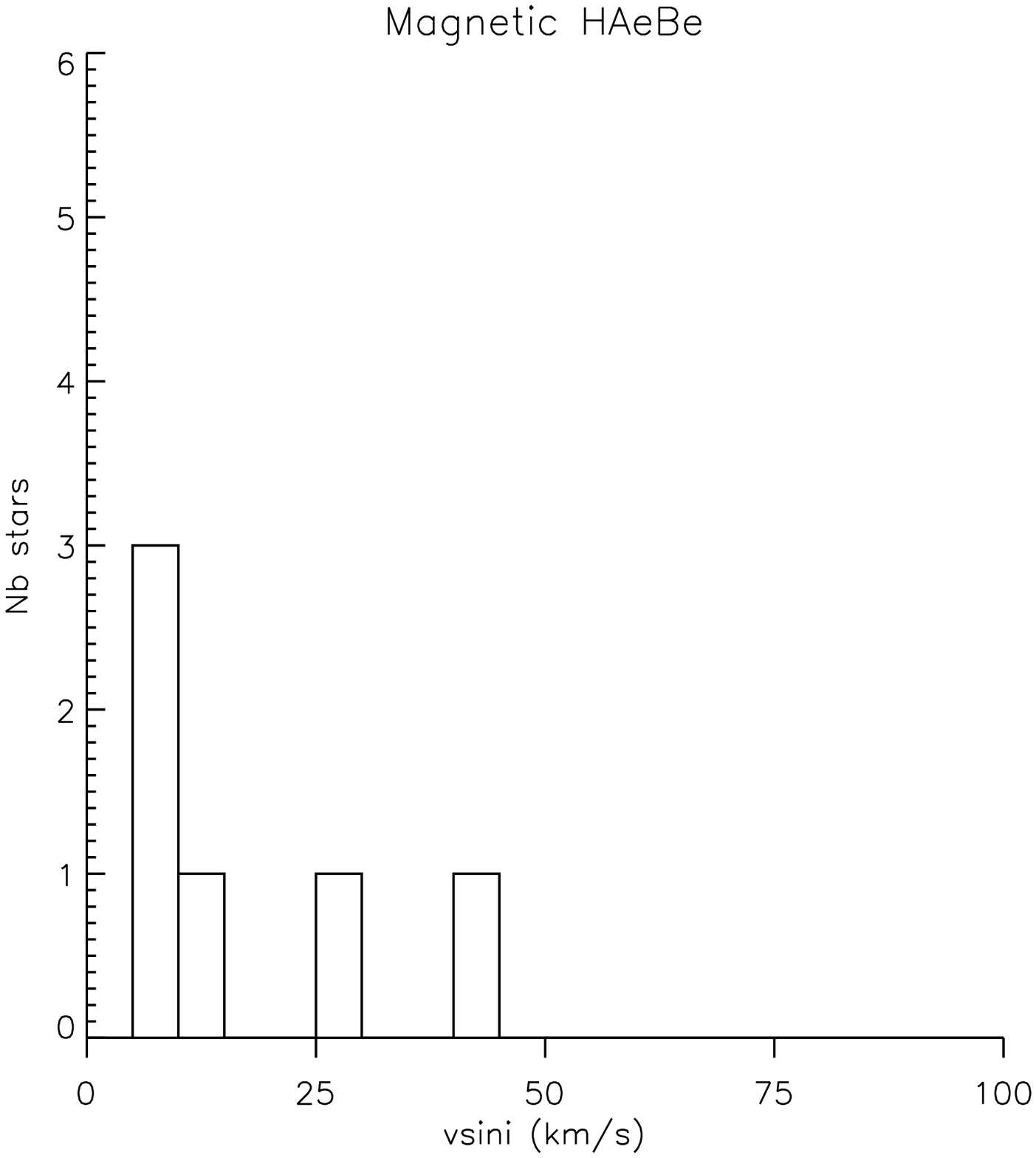}
\includegraphics[width=4cm]{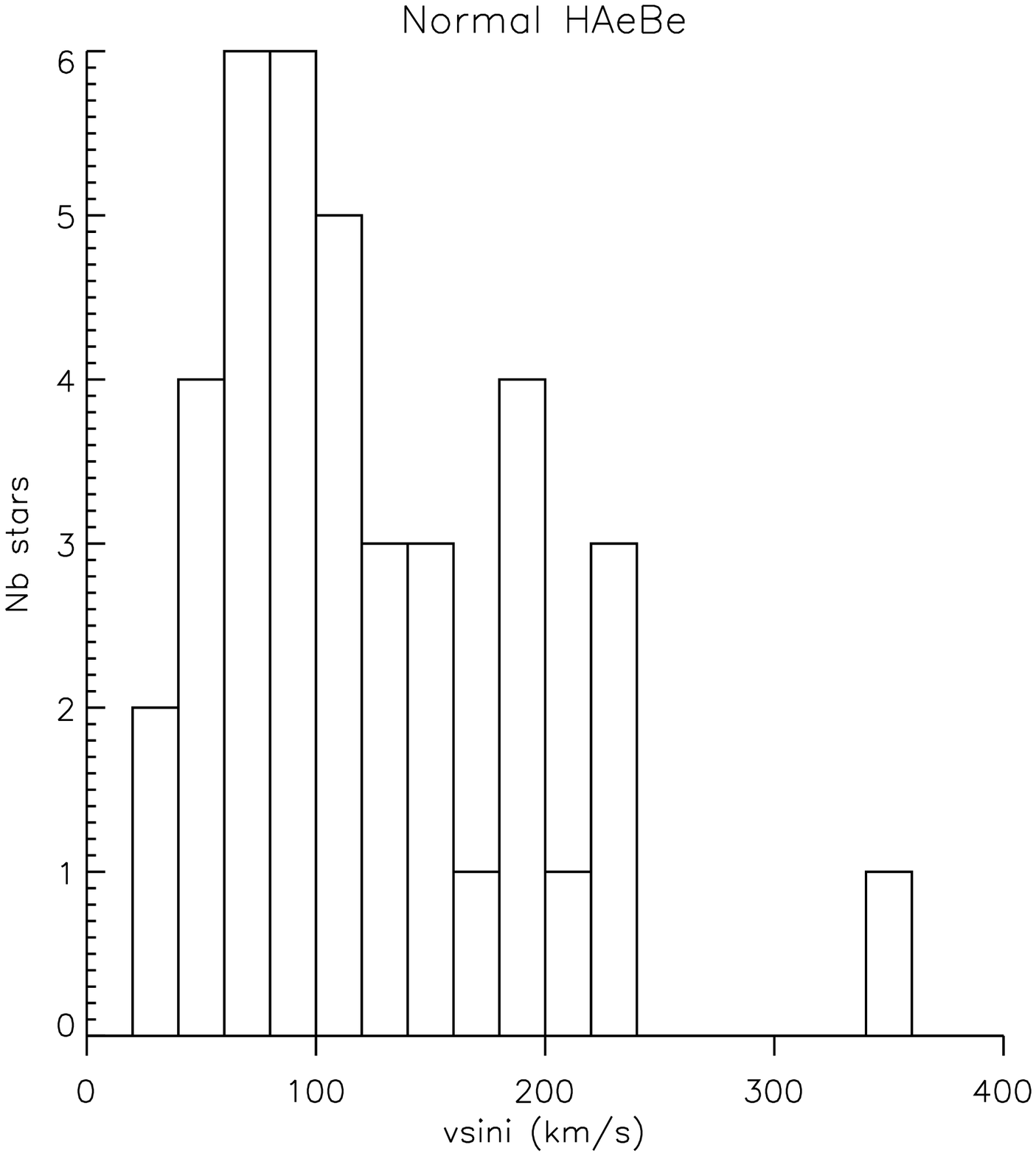}
\caption{$v\sin i$ histogram in number of stars, of the magnetic HAeBe stars ({\it left}) and of the normal HAeBe stars ({\it right}).}
\label{fig:rot1}
\end{figure}

In Fig. \ref{fig:rot1} are plotted the distributions of the $v\sin i$ of the magnetic HAeBe stars (left) and the normal HAeBe stars (right). Note that the binsize and the velocity scale are different in each case. We observe that the $v\sin i$ of the magnetic HAeBe stars are significantly lower than the $v\sin i$ of the normal ones: 4 magnetic HAeBe stars have a $v\sin i$ lower than 15~km.s$^{-1}$, while none of the normal HAeBe stars have a $v\sin i$ lower than 20~km.s$^{-1}$. Even though the inclination angles are not known and the true equatorial velocities of the stars are not known, it is clear that the magnetic HAeBe stars, as a population, are slower rotators than the normal HAeBe stars. We therefore conclude that the magnetic HAeBe stars seem to have been more braked than the normal ones.

\begin{figure}[t!]
\centering
\includegraphics[width=4cm]{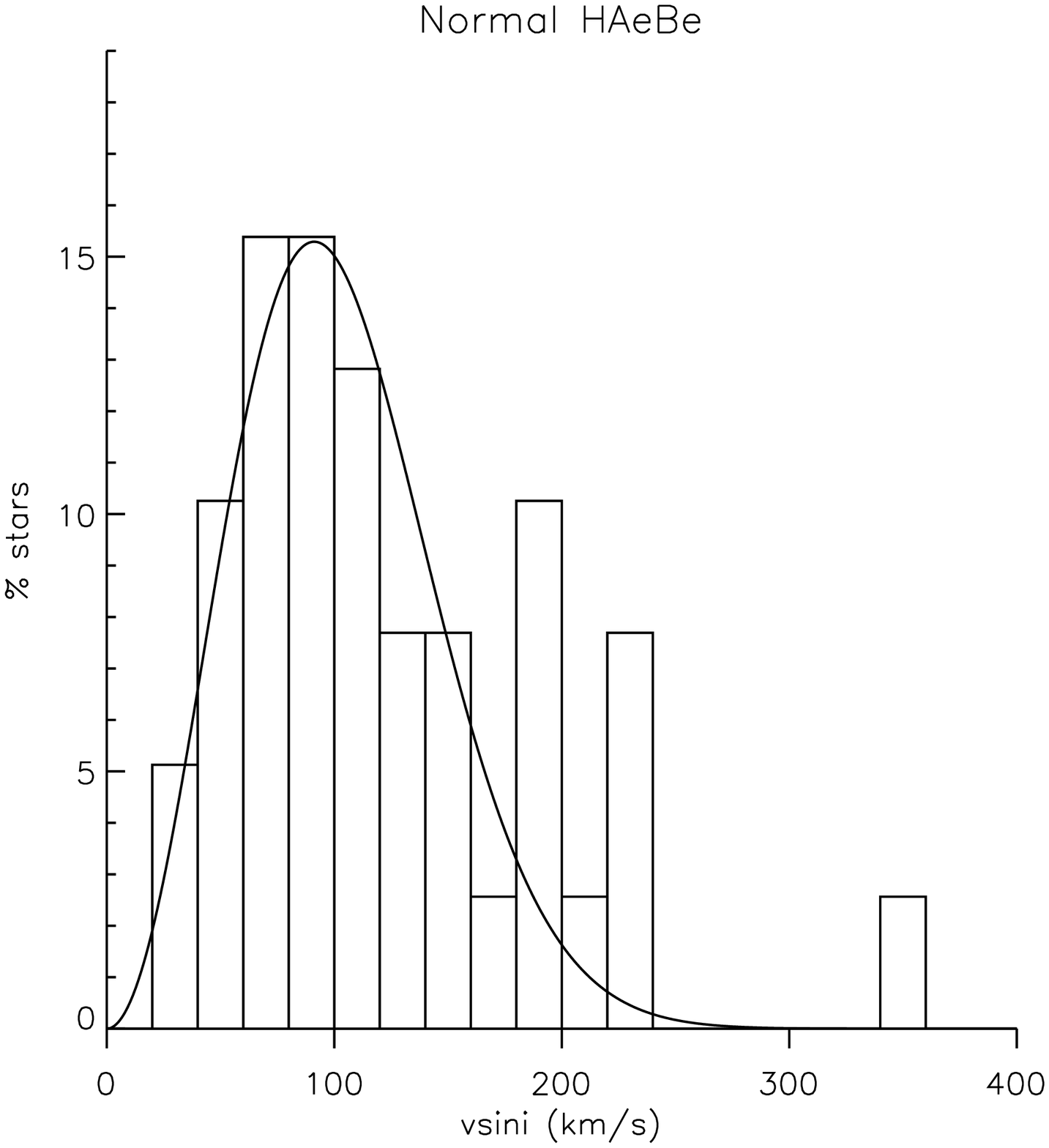}
\hfill
\includegraphics[width=4cm]{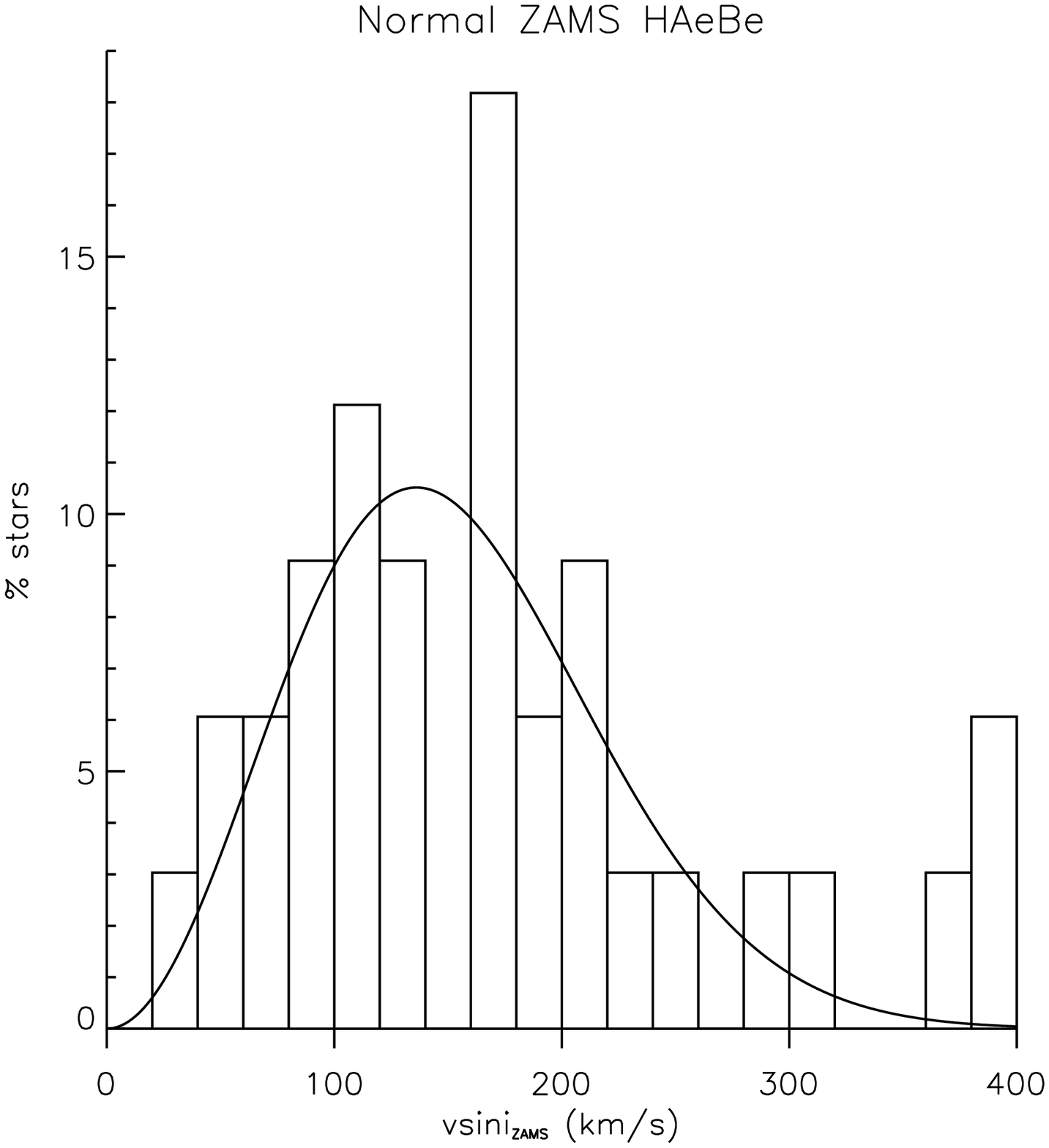}
\hfill
\includegraphics[width=4cm]{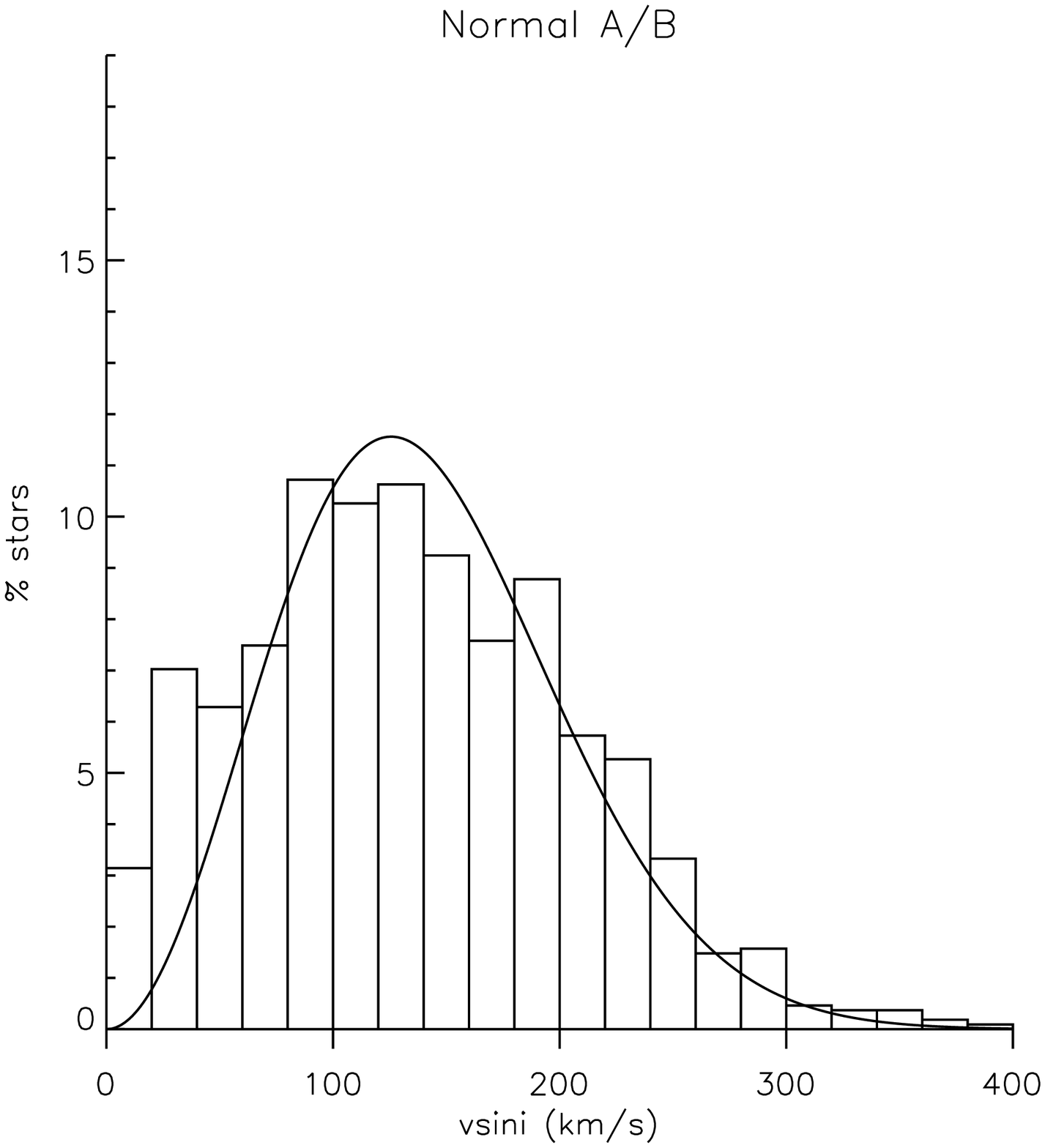}
\caption{$v\sin i$ histogram in percentage of stars of the normal HAeBe stars ({\it left}), of the normal HAeBe stars evolved to the ZAMS ({\it middle}) and of the normal A/B stars ({\it right}).}
\label{fig:rot2}
\end{figure}

PMS stars are still contracting towards the MS, therefore their radius is diminishing with time and their rotational velocity should increase towards the MS (assuming angular momentum conservation). In order to compare the rotation velocities of our normal HAeBe stars sample to those of the normal MS A/B stars, we compared the position in the HR diagram of each HAeBe star with CESAM evolutionary tracks. We inferred the mass, the actual radius and the age of the stars of our sample. We also inferred the radius that each star will have on the ZAMS, in order to compute their $v\sin i$ on the ZAMS. As the rotational velocity is slightly varying during the MS phase compared to the PMS phase, our distribution of $v\sin i$~(ZAMS) can be directly compared to the distribution of the $v\sin i$ of normal A/B stars. In Fig. \ref{fig:rot2} are plotted the distribution of the observed $v\sin i$ of normal HAeBe stars (left), their computed $v\sin i$ on the ZAMS (middle), and the observed $v\sin i$ of the normal A/B stars (right) from the catalogue of Royer et al. (2007). A Maxwellian function has been fitted to each distribution and is superimposed in full lines in Fig. \ref{fig:rot1}.

First we observe small but important differences between the actual $v\sin i$ and the ZAMS $v\sin i$ of HAeBe stars. The mode and the root mean square (rms) velocities of the Maxwellian function are respectively 91 km.s$^{-1}$ and 112 km.s$^{-1}$ for the actual $v\sin i$, and 136 km.s$^{-1}$ and 167 km.s$^{-1}$ for the ZAMS $v\sin i$. This slower and narrower $v\sin i$ distribution reflects the large range of ages and radii of the HAeBe sample. However these effects are small compared to what is expected during the entire PMS phase, according to the angular momentum conservation ($v\sin i$ should increase by a factor between 2 and 3, rather than 1.5 as observed). We explain this by the fact that most of our sample have less than 50\% of their PMS evolution remaining, during which the radius decrease is very small compared to the first part of the PMS phase (Alecian et al., in prep.). Our sample misses stars accomplishing the first half of their PMS phase, during which we cannot therefore conclude anything about the rotational evolution.

Secondly we observe a strong similarity between the predicted ZAMS $v\sin i$ of the HAeBe stars and those of the normal A/B stars: the modes and the rms velocities of the fitted maxwellian functions are respectively 136 km.s$^{-1}$ and 167 km.s$^{-1}$ for the ZAMS $v\sin i$ of the HAeBe sample, and 126 km.s$^{-1}$ and 154 km.s$^{-1}$ for the normal A/B stars. This suggests that the evolution of the PMS star from their actual age to the ZAMS is consistent with angular momentum conservation evolution.

%
\section{Herbig stars in young open clusters}

Motivated by these results we decided to go further and to observe HAeBe stars in young open clusters and associations. We selected Herbig members of three clusters: NGC 2244, NGC 2264 and NGC 6611, from the catalogues of Park \& Sung (2002), Park et al. (2000) and de Winter et al. (1997).

\begin{figure}[t]
\centering
\includegraphics[width=4cm, angle=90]{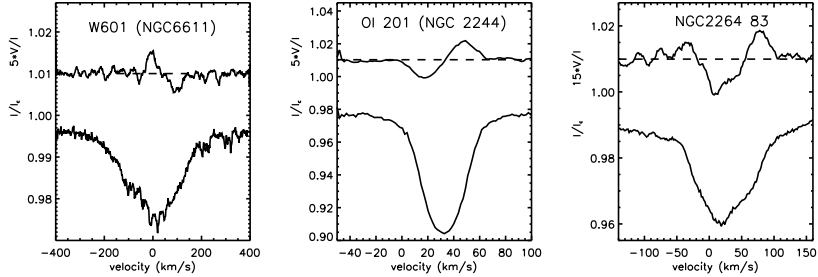}
\caption{Stokes $I$ (bottom) and Stokes $V$ (up) LSD profiles plotted for W601, OI~201 and NGC 2264 83 (Alecian et al. 2008b).}
\label{fig:autre}
\end{figure}

We observed 17 stars in NGC 2244, 27 stars in NGC 2264 and 17 stars in NGC 6611, which are plotted on the HR diagram in Fig. 1 (left, triangles). We detected magnetic fields in one star of each cluster: W601 (NGC~6611), OI~201 (NGC 2244) and NGC~2264~83 (Fig. \ref{fig:autre}).

The detection of W601 is very interesting, as this is our only detection of magnetic field in a fast rotator (${\rm v}\sin i\sim180$~km.s$^{-1}$). Furthermore, this star is very young ($<1\;$Myr) compared to OI~201 ($\sim2\;$Myr) and NGC~2264~83 ($\sim6\;$Myr). The two former have similar spectral type: B1.5 and B1 respectively, but W601 rotates with  ${\rm v}\sin i\sim180$~km.s$^{-1}$, while OI~201 rotates with only ${\rm v}\sin i\sim25$~km.s$^{-1}$ (Alecian et al. 2008b). These stars may provide us with clues about angular momentum evolution during the pre-main sequence phase.

%
\section{Conclusion}

Our spectropolarimetric study of Herbig Ae/Be stars among field and cluster stars has brought new strong arguments in favour of the fossil field hypothesis to explain the origin of the magnetic fields of the intermediate mass stars:
\begin{enumerate}
\item The projection of the magnetic field incidence of A/B stars onto the PMS phase predicts between 2 and 10 \% of magnetic PMS stars (Power et al., in prep.). This is consistent with our results: from our investigation, we fins that 7\% of HAeBe stars are magnetic.
\item The magnetic HAeBe stars host large-scale organised magnetic fields, similar to the Ap/Bp stars.
\item Using the actual radius and the predicted ZAMS radius of the field magnetic HAeBe stars and assuming magnetic flux conservation, we predict that their magnetic field intensities on the ZAMS will be between 0.4 and 4 kG (Table \ref{tab:fitv}). This is of the same order of magnitude as the magnetic intensities of the Ap/Bp stars.
\end{enumerate}
 Furthermore, in order to confront the fossil field hypothesis and the core dynamo hypothesis, we plotted in the HR diagram of Fig. \ref{fig:hr} (right panel) all the magnetic HAeBe stars, as well as the undetected field and cluster stars. We observe that 5 stars among the 8 magnetic HAeBe stars are in the {\bf totally radiative zone}. Although the error bars are large, one of these is confidently in the totally radiative zone, including its error bars on temperature and luminosity. On the other hand, all 8 magnetic HAeBe stars host the same type of magnetic fields: a large-scale organised magnetic field. We therefore conclude that presence of core convection does not appear to be responsible for the presence of magnetic fields in HAeBe stars. The magnetic fields of the intermediate mass stars are therefore {\bf very likely of fossil} origin.

The study of the rotation of field HAeBe stars leads to the conclusion that the magnetic HAeBe stars are more braked than the normal (non-magnetic and non-binary) ones. We also show that the evolution of the PMS stars from the HAeBe ages to the ZAMS is {\bf consistent with angular momentum conservation}. On the other hand our field HAeBe sample is too old to study the angular momentum evolution during the early stages of the PMS phase: most of our sample has already completed at least 40\% of their PMS track. Our HAeBe cluster sample contains very young PMS stars and will allow us to better understand stellar angular momentum evolution at all stages of the PMS phase.

\end{document}